\documentclass[pre]{revtex4}

\usepackage{amssymb}
\usepackage{amsmath}
\usepackage{graphics}
\usepackage{graphicx}
\usepackage{verbatim}
\usepackage{mathcomp}

\newcommand{\be}{\begin{equation}}
\newcommand{\ee}{\end{equation}}
\newcommand{\cone}{\textbigcircle $\hspace{-5mm}$ \begin{footnotesize} 1 \end{footnotesize}}
\newcommand{\ctwo}{\textbigcircle $\hspace{-5mm}$ \begin{footnotesize} 2 \end{footnotesize}}
\newcommand{\cfive}{\textbigcircle $\hspace{-5mm}$ \begin{footnotesize} 5 \end{footnotesize}}

\begin{document}

\title{The price of anarchy in basketball}
\date{\today}
\author{Brian Skinner}
\affiliation{School of Physics and Astronomy, University of Minnesota, Minneapolis, Minnesota 55455}

\begin{abstract}

Optimizing the performance of a basketball offense may be viewed as a network problem, wherein each play represents a ``pathway" through which the ball and players may move from origin (the in-bounds pass) to goal (the basket).  Effective field goal percentages from the resulting shot attempts can be used to characterize the efficiency of each pathway.  Inspired by recent discussions of the ``price of anarchy" in traffic networks, this paper makes a formal analogy between a basketball offense and a simplified traffic network.  The analysis suggests that there may be a significant difference between taking the highest-percentage shot each time down the court and playing the most efficient possible game.  There may also be an analogue of Braess's Paradox in basketball, such that removing a key player from a team can result in the improvement of the team's offensive efficiency.

\end{abstract} \maketitle

\section{Introduction}

In its essence, basketball is a network problem.  Each possession has a definite starting point (the sideline or baseline in-bounds pass) and a definite goal (putting the ball in the basket).  Further, each possession takes place through a particular ``pathway": the sequence of player movements and passes leading up to the shot attempt.  When a coach diagrams a play for his/her players, he/she is essentially instructing them to move the ball through a particular pathway in order to reach the goal.  If we think of a basketball offense as a network of possibilities for moving from in-bounds pass to goal, then we must imagine each pathway to have a different efficiency that depends in general on its frequency of use.  As an example, imagine a simple give-and-go play between the point guard and the power forward.  It could be expressed as a simplified pathway like this: 
\be 
(\textrm{point guard}) \stackrel{\rm{pass}}{ \longrightarrow} (\textrm{power forward}) \stackrel{\rm{pass}}{ \longrightarrow} (\textrm{point guard})\stackrel{\rm{shot}}{ \longrightarrow} (\textrm{basket}). \nonumber
\ee 
Each arrow represents a movement of the ball, beginning in the point guard's hands and ending in the basket.  The efficiency of this particular pathway (its likelihood of producing points) will be unique, and different from all other pathways such as the variation
\be
(\textrm{point guard}) \stackrel{\rm{pass}}{ \longrightarrow} (\textrm{power forward}) \stackrel{\rm{pass}}{ \longrightarrow} (\textrm{center})\stackrel{\rm{layup}}{ \longrightarrow} (\textrm{basket}). \nonumber
\ee 
The efficiency of a play should also depend on the extent to which it is used.  The give-and-go, for example, may be very effective when used seldomly, but if the team tries to run it every time down the court then the defense will learn to anticipate their opponents' moves and the offense will find much more modest success.  In principle, the full description of a basketball offense against a particular team would require every possible pathway to be diagrammed and assigned a unique function that describes its efficiency as a function of frequency of use.

At first sight, the network approach to basketball analysis seems hopeless.  Even if the complementary ``network" of defensive rotations and adjustments is not considered explicitly, there are still five interacting offensive players and an infinite set of potential plays and broken plays to be accounted for.  We should certainly not expect basketball to ever be a ``solved problem" (like Checkers or $5 \times 5$ Go).

The purpose of this article is to suggest that the network approach may nonetheless provide valuable insight into the operations and inefficiencies of a basketball offense.  In particular, the idea of a ``price of anarchy" may provide a useful definition for measuring the difference between a team's operating efficiency and its maximum possible production.  In this article I make a formal analogy between a basketball offense and a simplified traffic network.  By considering how optimal strategies are calculated and evaluated in the study of traffic patterns --- another ``hopelessly complicated" network problem --- we may gain insight into how such techniques can be applied to basketball analysis.

The following section introduces the concept of the price of anarchy and the method for calculating optimum strategies by discussing a simple model of traffic flow.  Section \ref{sec:onthecourt} discusses the price of anarchy as it pertains to basketball.  Two examples of simplified offensive networks are presented, and in each case the optimal strategy is calculated and contrasted with a ``short-sighted" strategy that is analogous to the Nash equilibrium in game theory.  A possible analogue of ``Braess's paradox" in basketball is also discussed, wherein the removal of a key player from a team may result in an improvement of the overall offensive efficiency.  The article concludes with some discussion of the limitations of the approach, and of possible applications to other sports.

\section{The price of anarchy on the highway} \label{sec:onthehighway}

The following argument is adapted from Tim Roughgarden's \textit{Selfish Routing and the Price of Anarchy} (2005), as described by Youn \textit{et. al.} (2008), and considers the simplest possible nontrivial network: two points A and B connected by a pair of parallel roads.

\begin{figure}
\centering
\includegraphics[width=0.45\textwidth]{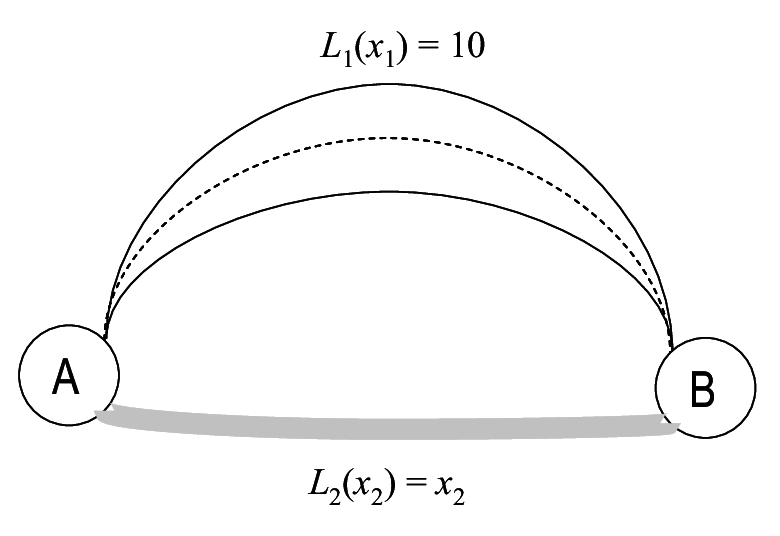}
\caption{Two paths by which traffic can travel from A to B: the long, wide highway (top) and the short, narrow alley (bottom).}
\label{fig:routes}
\end{figure}

Imagine that ten cars have to get from point A to point B, and that connecting A and B are two roads, as shown in Fig$.$ \ref{fig:routes}.  The upper road (labeled $1$) is a wide highway.  It takes ten minutes to traverse regardless of the number of cars on it.  In other words, the duration $L_1$ of a driver's commute on the highway as a function of the number of cars $x_1$ on it is just a constant, $L_1(x_1) = 10$.  The lower road (labeled $2$) is a narrow alley.  It is more direct but cannot hold much traffic without becoming significantly slower.  Say that the duration of a commute along the alley depends on the number of cars $x_2$ that take it as $L_2(x_2) = x_2$.  That is, if only one car takes the alley, then that car has a 1-minute commute.  If two cars take the alley, then they each have a 2-minute commute, and so on.

Evidently, there is no reason for anyone to take the highway.  At worst, driving through the alley will take ten minutes, while the highway always takes ten minutes.  It is therefore in the best interest of all ten drivers to take the alley.  This situation results in an average commute time $\langle L \rangle_{NE} = 10$, and is called the ``Nash equilibrium".  In general, the Nash equilibrium is the state where no individual can improve his situation by making a different choice.  In this particular case, it is the state where each driver has the same length of commute.

However, for the drivers in this example, the Nash equilibrium is not actually the best possible outcome.  For arbitrary usage levels $x_1, x_2$ of the two roads, the average commute length is
\be 
\langle L \rangle = \frac{x_1 L_1(x_1) + x_2 L_2(x_2)}{10} = x_1 + x_2^2/10 .
\ee 
Since there are ten cars total, $x_1$ and $x_2$ satisfy $x_1 + x_2 = 10$, and by substitution we have
\be 
\langle L \rangle = x_1 + (10 - x_1)^2/10 .
\ee 
The minimum value of $\langle L \rangle$ satisfies $d\langle L \rangle/dx_1 = 0$, which gives
\begin{eqnarray}
x_{1, opt} = x_{2,opt} = 5, \\
\langle L \rangle_{opt} = 7.5. \nonumber
\end{eqnarray}
This is an important conclusion: the global optimum solution is the one where the roads are used equally, even though they are not of equal quality.  The optimum scenario has $5$ cars getting 10-minute commutes and $5$ others getting 5-minute commutes, while at the Nash equilibrium all $10$ cars had 10-minute commutes.  It should be noted that this optimum is not stable, in a social sense.  Each of the $5$ people driving on the highway has an incentive to switch to the alley, whereby he/she could reduce his/her driving time from 10 minutes to 6.  So if the global optimum is to be maintained, there must be some kind of structure enforced upon the drivers.  That is, the drivers must be ``coached" into the global optimum, because it is against their immediate best interest.  The difference between the ``selfish" Nash equilibrium and the real optimum is called the ``price of anarchy."  Here it is $\langle L \rangle_{NE} - \langle L \rangle_{opt} = 2.5$ minutes, or 33\%.  

If more traffic complicated networks are considered, then the discrepancy between the Nash equilibrium and the global optimum can allow for counterintuitive consequences associated with opening and closing roads.  In terms of the global optimum, closing a road must always have a negative effect; it removes potential options whereby traffic may flow.  However, closing a road can have the effect of greatly disturbing the Nash equilibrium, and perhaps pushing it closer toward the global optimum distribution.  Since real-world traffic follows something much closer to the Nash equilibrium than the global optimum, it is fully possible that closing down a road can \textit{improve} traffic.  This phenomenon is called ``Braess's Paradox", and it has apparently been observed in a number of major cities like New York, San Francisco, and Stuttgart, where the closing of a major road was accompanied by almost immediate improvements in traffic flow (see Youn \textit{et. al.}, 2008, and references therein).

\section{The price of anarchy on the court} \label{sec:onthecourt}

At this point the analogy to basketball is probably clear.  On the basketball court, possessions are like cars.  Each one starts at point A (the in-bounds) and attempts to travel to point B (the basket).  Different plays are like different roads: each one has a different efficiency that will generally decrease the more it is used.  In principle, all of the methodologies and ``paradoxes" associated with traffic patterns should be applicable to basketball as well.

It may seem like common sense to say that every play in a basketball game should be run in the way that gives it the highest probability of success.  But, as this section will demonstrate, such a strategy is like the ``selfish" Nash equilibrium in the previous section, and it is not necessarily the one the leads to the highest overall efficiency.  A team that contents itself with making the highest-percentage play each time down the court and does not consider the game-wide implications of its strategy can pay a significant price of anarchy for its ``short-sighted" approach.

\subsection{A simplified offensive network}

\begin{figure}
\centering
\includegraphics[width=0.5\textwidth]{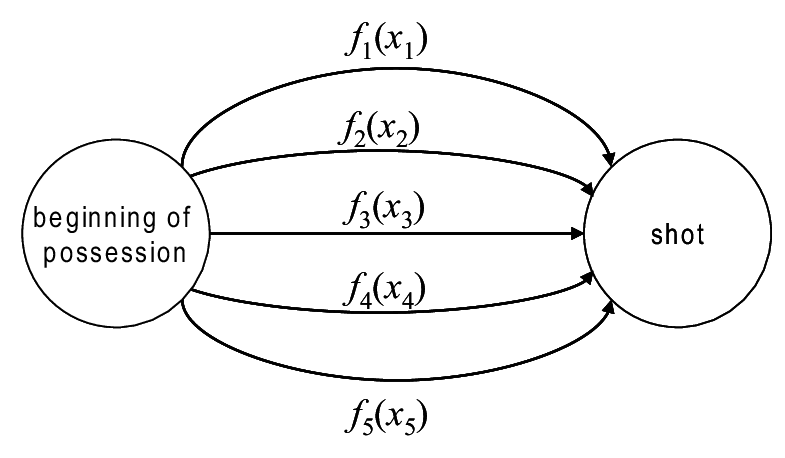}
\caption{Five paths by which a team may aquire points.  Each path represents a shot attempt by one of the team's five players.}
\label{fig:shotroutes}
\end{figure}

Motivated by the traffic picture of Fig$.$ \ref{fig:routes}, we can try to draw the simplest possible network for a basketball offense: it should look something like Fig$.$ \ref{fig:shotroutes}.  Each line connecting the beginning of the possession to the shot attempt represents a different player, who is treated as a possible path by which points may come.  Each player (labeled $i = 1, 2, ..., 5$) has a particular scoring efficiency $f_i$ that depends on his frequency of use $x_i$.  

Of course, Fig$.$ \ref{fig:shotroutes} is a drastic and brutish over-simplification of the beautiful complexities of a real offense: it assumes that the only thing relevant for the efficiency of a shot is who takes it.  In reality, the efficiency of a shot should depend on the play leading up to it, but this simplified model may nonetheless be instructive as a first attempt.

\subsection{Skill curves}

Before going any further, we should carefully define what is meant by ``efficiency of a shot".  The most useful definition for efficiency is something like ``probability that a shot produces points".  While field goal percentage would be a close approximation, a slightly more advanced statistic has already been invented for this concept.  It is called ``true shooting percentage", or TS\%, and is best thought of as a field goal percentage adjusted for free throws and three-point shots, or $1/2 \times (\textrm{points per shot})$ (see Kubatko \textit{et. al.}, 2007).  TS\% is defined by the formula
\be 
\textrm{TS\%} = \frac{\frac12 (\textrm{points scored})} { (\textrm{field goal attempts}) + 0.44 \times (\textrm{free throw attempts}) }.
\ee 
The efficiency function $f_i(x_i)$ of a player can therefore be defined as the player's TS\% as a function of the fraction of the team's shots $x_i$ that he takes while on the court.  

Such a function is notoriously difficult to extract from commonly-recorded statistics, and in fact the validity of prescribing a set function to a player's efficiency at all is contested.  It isn't hard to believe, however, that a player's efficiency should decline the more he is used.  As an extreme example, any player who made up his mind to shoot \textit{every time} down the court would surely be less efficient than if he took a more judicious $1/5$ of the team's shots.

In \textit{Basketball on Paper} (2004), Dean Oliver envisioned such an inverse relationship between offensive usage and offensive efficiency.  He dubbed it a ``skill curve", although he used somewhat different definitions of ``usage" and ``efficiency" than are employed here.  The skill curve remains a highly theoretical object that has never been extracted from data in a generally accepted way.  This difficulty arises because the volume of a player's shots usually doesn't vary over a sufficiently wide range to make a good characterization.  A great player will not spend much time taking less than 10\% of his team's shots, and a mediocre player will not spend much time taking more than 50\% of his team's shots.  So it is hard to reliably judge how much a player's efficiency declines with usage.  There is also the difficulty that a skill curve is meant to be defined \textit{against a given defense}; a player's efficiency $f(x)$ may be noticeably different against team A than against team B, depending on their respective caliber of defense.

\begin{figure}
\centering
\includegraphics[width=.5\textwidth]{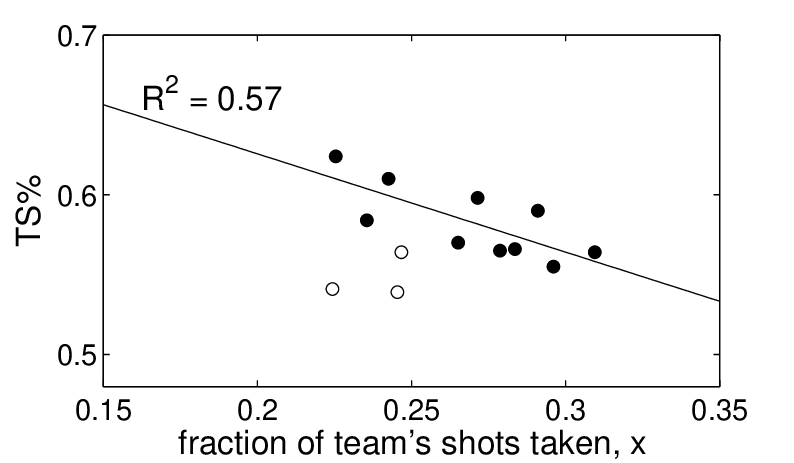}
\caption{Estimated ``skill curve" for NBA shooting guard Ray Allen.  Each data point represents the average performance over a given season of his professional career.  The solid line is a linear fit, shown with its $R^2$ correlation value.  Open circles represent the first three seasons of his career, and are not included in the fit line.}
\label{fig:ray_allen-skill_curve}
\end{figure}

Nonetheless, something approximating a skill curve can be estimated by examining season averages for TS\% and shot volume among players whose offensive load has varied significantly over the course of their playing career.  This approach assumes that the league as a whole maintains a similar quality of defense from year to year, so that the skill curve is defined in terms of the average defender that the player is likely to encounter.  As an example, I consider the season averages for current Boston Celtics guard Ray Allen.  Fig$.$ \ref{fig:ray_allen-skill_curve} shows a plot of Allen's TS\% as a function of the fraction of his team's shots $x$ that he took while on the court; the solid line is a linear best-fit.  The first three seasons of his career (open circles) were not included in the fit lines, as this was considered a ``learning curve" period wherein Allen was still improving significantly.  The fraction $x$ was calculated as
\be 
x = \frac{(\textrm{player shots/game})}{(\textrm{team shots/game})} \times  \frac{(\textrm{48 minutes/game})}{(\textrm{player minutes/game})}.
\ee 
All data was provided by http://www.basketball-reference.com/.  

Given the data in Fig$.$ \ref{fig:ray_allen-skill_curve}, the existence of a negative relationship between shot volume and shot efficiency seems at least plausible.  It is reasonable to assume that, over the relevant range of shot fraction $x$, the function $f(x)$ is linear.  Further analysis will use this assumption.

\subsection{Equilibrium versus optimum} \label{sec:formalism}

Suppose that a team of five players is aware of each other's skill curves.  How should they distribute their shots in order to achieve optimal efficiency?  The most obvious strategy is to always give the ball to the player with the highest percentage chance of making the shot.  That is, if player A is shooting with a higher TS\% than player B, this strategy calls for the team to apportion more shots to player A, which process  continues until player A's efficiency declines to the level of player B's (or player B's efficiency rises to the level of player A's).  The result is to establish a kind of equilibrium among all five players, such that  all five shoot the same percentage (or don't shoot at all).  In other words, the strategy of looking for the highest percentage option each time down the court results in an equilibrium defined by 
\be 
f_1(x_1) = f_2(x_2) = ... = f_5(x_5),  \label{eq:fair}
\ee
and subject to the constraint
\be 
x_1 + x_2 + ... + x_5 = 1. \label{eq:normalization}
\ee

It is probably incorrect to call this strategy a ``Nash equilibrium", since the term is meant to describe a fixed point among competing players and not among teammates with a single common objective.  However, the analogy with the Nash equilibrium in the traffic example is straightforward.  In traffic, the Nash equilibrium was reached when each driver took the path that was in his/her best interest.  The equilibrium of equation (\ref{eq:fair}) is reached when each \emph{possession} follows the path that is in \emph{its} best interest.  Such a situation could be called a ``short-sighted equilibrium"; here I will continue to use the term ``Nash equilibrium" in order to maintain the analogy with traffic networks.

It is the surprising result of the example in section \ref{sec:onthehighway} that the Nash equilibrium is not necessarily the team's best strategy.  The true best strategy will be the one that optimizes the team's overall efficiency $F$, defined as
\be 
F = x_1 f(x_1) + x_2 f(x_2) + ... + x_5 f(x_5).
\ee
The optimization condition is
\be 
\frac{dF}{d x_1} = \frac{dF}{d x_2} = ... = \frac{dF}{d x_5} = 0, \label{eq:opt}
\ee 
and is again subject to the ``normalization" of equation (\ref{eq:normalization}).  If the team's efficiency when it follows the optimum strategy of equation (\ref{eq:opt}) is denoted $F_{opt}$, and its efficiency when it follows the Nash equilibrium is denoted $F_{NE}$, then the price of anarchy for the offense is defined as $F_{opt} - F_{NE}$.

Under the assumption that each player's skill curve $f_i(x_i)$ is a linear function
\be 
f_i(x_i) = \alpha_i - \beta_i x_i,
\ee 
where $\alpha _i$ and $\beta_i$ are positive numbers such that $\alpha_i \le 1$ and $\beta_i \le \alpha_i$, the optimum strategy can be calculated exactly by using the method of Lagrange multipliers for constrainted optimization.  Here $\alpha_i$ represents the highest possible TS\% that player $i$ can shoot and $\beta_i$ represents the rate at which the player's efficiency declines with increased usage.

Specifically, the optimum values of $x_i$ satisfy
\be 
\frac{\partial}{\partial x_i} \left( F + \lambda (\sum_{i = 1}^5 x_i - 1) \right) = 0, \label{eq:lagrange}
\ee
where $\lambda$ is a constant called the Lagrange multiplier.  The solution of equation (\ref{eq:lagrange}) is straightforward:
\be 
x_{i,opt} = \frac{\alpha_i + \lambda}{2 \beta_i}, \label{eq:optx}
\ee
where
\be 
\lambda = \frac{ 2 - \sum_{i = 1}^5 \alpha_i/\beta_i }{\sum_{i = 1}^5 1/\beta_i }. \label{eq:Lmultiplier}
\ee
The corresponding optimum efficiency
\be 
F_{opt} = \sum_{i = 1}^5 x_{i,opt} f_i(x_{i,opt}) = 
\sum_{i = 1}^5 \frac{\alpha_i^2 - \lambda^2}{4 \beta_i} \label{eq:Fopt}.
\ee

It should be noted that in some (fairly unrealistic) situations, equations (\ref{eq:optx}) and (\ref{eq:Lmultiplier}) may give a negative value for the shot fraction $x_{j,opt}$ of some player $j$.  Such a solution, of course, is spurious, and implies that in the optimum solution player $j$ is not apportioned any shots at all.  The true optimum in this instance can be found by removing player $j$ from the offensive network and performing the relevant sums in equations (\ref{eq:optx}) -- (\ref{eq:Fopt}) over the remaining four players.

\subsection{Example: Ray Allen as the efficient alley}

To illustrate the significance of the optimum strategy, and the extent to which it can differ from the ``short-sighted" Nash equilbrium, we can consider a simple example analogous to the one of Fig$.$ \ref{fig:routes}.  Suppose that Ray Allen is on a team with four ``lesser" players, who generally have a lower TS\% than he does.  The plot in Fig$.$ \ref{fig:ray_allen-skill_curve} suggests that Ray Allen's TS\% is
\be 
f_{RA}(x) \approx 0.75 - 0.62 x.
\ee 
Imagine now that Allen's teammates shoot with a constant efficiency $f = 0.5$, regardless of their shot volume (which will never be larger than 25\% each, anyway).  In this scenario, Ray Allen is like the ``alley", which provides a more efficient path to the goal but declines with usage, while Allen's teammates are like the ``highway", which is less efficient but does not decline in efficiency with usage.

The Nash equilibrium strategy would require that Ray Allen continue shooting until he is only as efficient as his teammates: $f_{RA}(x_{NE}) = 0.5$.  Consequently, Ray Allen would take $x_{NE} \approx 0.40$ of his team's shots and the team would have a TS\% of $F_{NE} = 0.5$.

\begin{figure}
\centering
\includegraphics[width=.5\textwidth]{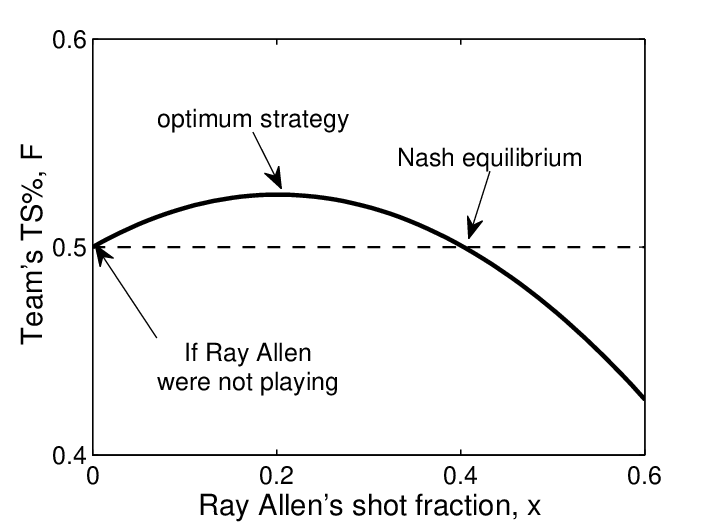}
\caption{The TS\% of Ray Allen's team as a function of the fraction of the team's shots he takes, as given by Eq$.$ (\ref{eq:RAteam}).}
\label{fig:RAteam}
\end{figure}

In this extreme case, it is probably clear that the Nash equilibrium is not the optimal strategy.  A team with Ray Allen, who is an exceptional player, should not shoot only as well as his less-than-exceptional teammates.  More generally, if Ray Allen shoots some fraction $x$ of his team's shots, then the team's TS\% will be
\be 
F = x f_{RA}(x) + (1 - x) \times 0.5 . \label{eq:RAteam}
\ee 
The team's efficiency is optimized when $dF/dx = 0$, which gives $x_{opt} = 0.202$ and a corresponding TS\% of $F_{opt} = 0.525$.  The price of anarchy is therefore $F_{opt} - F_{NE} = 0.025$. Fig$.$ \ref{fig:RAteam} shows the function $F(x)$, and highlights the different equilibrium points.

This result is fairly surprising.  It suggests that the team's optimum strategy is for Ray Allen to shoot almost exactly the same fraction of shots as his teammates --- about 20\%.  In such a scenario Allen would have a TS\% of $0.625$, far above the rest of his teammates who are shooting with a TS\% of $0.5$.  It would seem to be obvious that Ray Allen should be getting more shots than his teammates.  But the result of limiting Allen's shots, and thereby keeping the defense from focusing too intently on him, pays off.  In this case it improves the team's TS\% by 2.5\%.  For reference, in the 2008-2009 NBA season the standard deviation among teams in TS\% was only 1.6\%, so a difference of 2.5\% may be quite significant.  To achieve it, the team just needs to consciously choose not to have Ray Allen shoot on $\sim 80\%$ of possessions, even when he is their best option.

\subsection{Braess's Paradox and the ``Ewing Theory"}

An interesting implication of the model in the previous subsection is that, if Ray Allen's team is playing by the Nash equilibrium, then it doesn't make a difference whether he is in the game or not.  When Ray Allen plays, the Nash equilibrium settles at a team TS\% of $0.5$.  But the team would shoot this same percentage even if Ray Allen did not take a single shot.  So this hypothetical team, if playing by the Nash equilibrium and not with the optimal strategy, would lose nothing if Ray Allen sat out.

Of course, a real NBA offense is much more complicated than the diagrams of Fig$.$ \ref{fig:routes} or Fig$.$ \ref{fig:shotroutes}.  But that isn't necessarily a reason to dismiss the conclusion.  In traffic networks, it is entirely possible that removing a road --- even an efficient one --- can cause traffic to improve by pushing the Nash equilibrium closer to the true optimum.  In basketball, it seems plausible that losing an efficient player could push the team closer to its true optimum and thereby increase the team's efficiency.  In traffic, this counterintuitive effect is called ``Braess's Paradox".  In sports, \textit{ESPN.com} sportswriter/humorist Bill Simmons (2001) called it the ``Ewing Theory".  The idea that a team could improve after losing one of its best players may in fact have a network-based justification, and not just a psychological one.

Following the example of traffic networks which demonstrate Braess's Paradox (Youn, 2008), it isn't difficult to construct an offensive network which exhibits a similar phenemonon.  That is, one can easily imagine a network of plays whose Nash equilibrium will improve when a key player is removed.

\begin{figure}
\centering
\includegraphics[width=.5\textwidth]{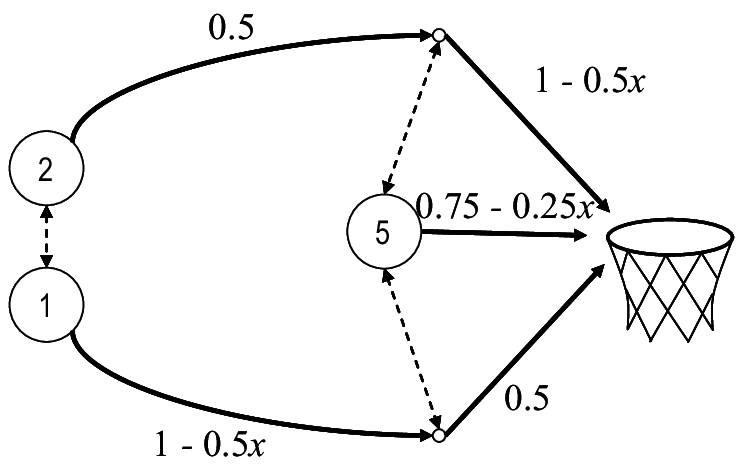}
\caption{A dribble penetration play, which starts with either the point guard \textcircled1 or the shooting guard \textcircled2 driving to the basket, and ends with a shot by one of the guards or the center \textcircled5 .}
\label{fig:drive_and_kick}
\end{figure}

Consider, for example, the network diagrammed in Fig$.$ \ref{fig:drive_and_kick}, which is designed to produce a layup for either the point guard (labeled \cone), the shooting guard (\ctwo), or the center (\cfive).  The play starts with the ball in the hands of either  \cone$ $ or \ctwo, who drives from the top of the key toward the basket.  When he arrives at the low post, the driver has two options: he can put up a shot at the basket or he can shovel the ball to \cfive.  If \cfive  $ $ receives the ball, he also has two options: he can take a shot or he can pass the ball to a cutting \cone $ $ or \ctwo --- whichever guard it was that did not make the initial drive to the basket --- for the layup.  Fig$.$ \ref{fig:drive_and_kick} shows this network of possible ball movements.  Solid lines indicate a drive or shot attempt, and dashed lines represent passes.

Next to each solid line is a formula that is meant to show the effectiveness of that step of the play, \textit{i.e.} the probability that the given step will be completed successfully.  Here, $x$ stands for the fraction of the time that the given step is used; $x = 1$ if the ball always proceeds through this route and $x = 0$ if the route is never used.  The choice of efficiencies here is arbitrary, but the situation can be imagined like this: \cone $ $ is a quick and capable ball handler, but is mediocre at finishing around the rim.  \ctwo $ $ is not particularly good at driving the ball, but is significantly better at finishing around the basket.  The center \cfive $ $ is somewhere in the middle as a shooter: not as good as \ctwo, but better than \cone.  For simplicity, it is assumed that all passes are successful every time.

The Nash equilibrium can be easily calculated for this network once the following observations are made:

\begin{enumerate}
\item \ctwo $ $ is never more successful at driving the ball than \cone.  Even at his very worst, \cone $ $ is successful on 50\% of his drives while \ctwo $ $ is always 50\% successful.

\item Similarly, \cone $ $ is never more successful at shooting than \ctwo $ $ or \cfive .  At their very worst, \ctwo $ $ and \cfive $ $ are 50\% effective at finishing at the rim while \cone $ $ always shoots 50\%.
\end{enumerate}

By the Nash equilibrium strategy, where the effectiveness of each individual possession is optimized, there is apparently no reason for \ctwo $ $ to ever drive or for \cone $ $ to ever shoot.  That is, \cone $ $ should drive the ball every time and then pass to the center.  The only remaining question is how often \cfive $ $ should attempt the layup himself and how often he should pass to \ctwo $ $ for the finish.

In the Nash equilibrium strategy, this question is also easily answered by equating the shooting percentages of \cfive $ $ and \ctwo.  The result is for \cfive $ $ to take the shot $1/3$ of the time and \ctwo $ $ to take the shot $2/3$ of the time.  Both players make the layup $2/3$ of the time.  The overall efficiency of the play is therefore $0.5 \times (2/3) = 0.33$.

\begin{figure}
\centering
\includegraphics[width=.5\textwidth]{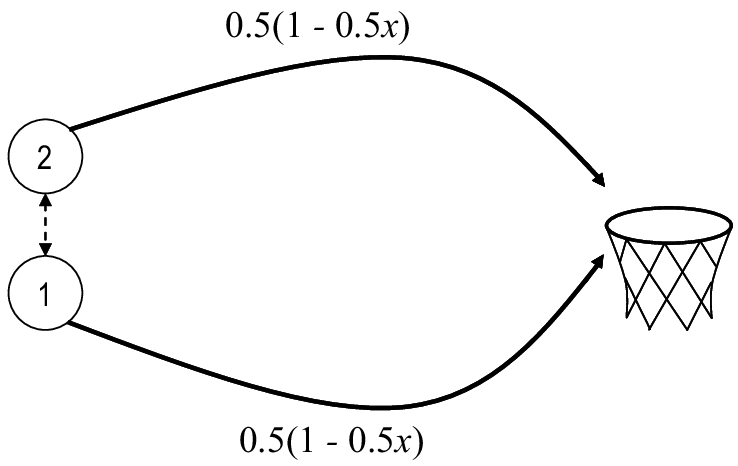}
\caption{The same dribble penetration play as in Fig$.$ \ref{fig:drive_and_kick}, but with the center \textcircled5 removed from the play.}
\label{fig:nocenter}
\end{figure}

We can now consider how the play might look if \cfive $ $ is removed from the offense --- perhaps the original player is lost to injury and a replacement is introduced who does not get involved in the offense, but merely stands off to the side when the play is run.  In this situation, all passes to or from \cfive $ $ are eliminated, and the only options for the play are a drive and shot by \cone $ $ or a drive and shot by \ctwo.  So the offensive network is simplified to the one depicted in Fig$.$ \ref{fig:nocenter}.

In this case, the two guards \cone $ $ and \ctwo $ $ constitute equivalently effective options for the play --- \cone $ $ is good at the drive and worse at the shot while \ctwo $ $ is good at the shot and worse at the drive, so that their overall efficiencies in running the play from the top of the key are equal.  As a consequence, the guards \cone $ $ and \ctwo $ $ split their shot attempts 50/50, and each scores with an efficiency $0.5 \times (1 - 0.5 \times 0.5) = 0.375$.  This is the same as the overall efficiency of the play, and, surprisingly, it is higher than the $0.33$ efficiency that resulted when \cfive $ $ was involved.

This counterintuitive result is completely analogous to Braess's Paradox in traffic networks: an important component of the network is removed, but the overall efficiency improves.  It is worth noting that in the original play, run at the Nash equilibrium, the center \cfive $ $ was an absolutely key contributor.  He touched the ball every time the play was run and shot $1/3$ of the time with a 67\% success rate.  A team that found its success improved after losing such a player would likely be extremely surprised.

Of course, this result really highlights the sub-optimality of the ``short-sighted" Nash equilibrium solution.  The true optimum performance for this offense can be found not by considering the best way to run each individual play, but by considering every option simultaneously and optimizing the total performance with respect to the frequency that each option is run.  Following the method outlined in section \ref{sec:formalism}, it is easy to show that the optimum strategy involves \cone $ $ driving half the time and \ctwo $ $ driving half the time.  As for the shot attempt, \ctwo $ $  and \cfive $ $ should split the shot attempt 50/50, even though in this case they will shoot very different percentages.  The resulting overall efficiency of the play is $0.43$, well above the Nash equilibrium solution $0.33$ or the efficiency $0.375$ that resulted when \cfive $ $ was removed.  

In other words, there is a significant price of anarchy for this offensive set.  When run with the best possible strategy, the play can be successful $43\%$ of the time.  But when the team only considers one possession at a time, the play works only $33\%$ of the time.  The apparent improvement of the team upon losing its center, and, indeed, any instance of the ``Ewing Theory", is a consequence of this short-sightedness.

\section{Conclusions}

The statements of this paper are intended to constitute a plausibility argument for the usefulness of network approaches and ``price of anarchy" analyses in basketball.  They should not be taken as proof of any particular phemomenon in basketball, and are not supported by any hard statistical evidence.  What's more, the defense has been treated only implicitly in this approach, and was not given the status of a ``player in the offensive network".  In principle, a defense should be able to choose to focus on a given player to an arbitrary degree, thus shifting that player's skill curve up or down to an arbitrary extent.  The assumption of using a set efficiency function is that the defense \emph{always} responds in a particular way to a player's increased usage.  It may be that the full incorporation of a defense's ``right to choose" will weaken the major conclusions of this paper.

Nonetheless, the idea of a difference between Nash-like equilibrium states and a global optimum strategy is enticing.  If a team could be fully aware of the efficiencies of all of its plays and players (its ``skill curves"), then it could easily determine the exact best way to allocate shots, passes, drives, \textit{etc$.$} and could quantify the difference between its current operating efficiency and its maximum potential.  Unfortunately, if the methods advocated in this paper are to be given any real predictive power, we need a way of reliably estimating a player's skill curve.  The rude approach taken in this paper is limited only to players with many years' worth of accumulated data, and whose offensive responsibilities have varied of the years.  The development of a trustworthy method for extracting skill curves from single-season data, or even from controlled experiments, would be a big step toward verifying the significance of the arguments presented here.

As a final comment, it should be noted that the points made in this paper are perhaps not novel at all from a coaching standpoint.  A basketball coach calling a play for his team is probably aware that he/she should not always simply call the play with the highest probability of success.  Rather, a coach may decide to ``save" the team's best plays for later, with the purpose of retaining their high efficiency against the defense.  In fact, this kind of reasoning is almost certainly in the minds of coaches and players in other sports.  In baseball, a pitcher will often decide not to throw his best pitch --- even if it means a higher chance of giving up a hit in the short term --- so that the batter is less prepared when he \textit{does} use his ``best stuff".  In American football, every professional team produces more yards with the average passing play than with the average running play (see Alamar, 2006).  Nonetheless, teams continue to employ less efficient running plays as a means of ``keeping the defense honest". That is, the immediate success of a given play is sacrificed in order to maintain the high efficiency of a team's offense as a whole.  This is exactly the type of reasoning advocated in this paper.

If such thinking is indeed already in the minds of coaches and players, then it should probably be in the minds of those who do quantitative analysis of sports as well.  It is my hope that the introduction of ``price of anarchy" concepts will constitute a small step towards formalizing this kind of reasoning, and in bringing the analysis of sports closer in line with the playing and coaching of sports.


\end{document}